# تحلیل پایداری و پایدارسازی سیستم‌های فازی تاکاگی-سوگنو مثبت بازه‌ای با استفاده از بهینه‌سازی محدب

الهام احمدی[1]، جعفر زارعی[2]

[1] دانشجوی کارشناسی ارشد، دانشگاه صنعتی شیراز، e.ahmadi@sutech.ac.ir

[2] دانشیار، دانشگاه صنعتی شیراز، zarei@sutech.ac.ir

**چکیده –** در این مقاله مساله پایداری و پایدارسازی سیستم‌های غیرخطی مثبت، توصیف شده با مدل فازی تاکاگی-سوگنو گسسته-زمان مورد مطالعه قرار می‌گیرد. رویکرد پیشنهادی بر مبنای تابع لیاپانوف خطی شبه مثبت و کنترل‌کننده توزیع شده موازی می‌باشد. شرایط لازم و کافی برای وجود کنترل‌کننده فیدبک حالت، که مثبت و پایدار بودن سیستم حلقه بسته را تضمین کند، به صورت برنامه‌ریزی خطی، با رویکردی جدید به دست می‌آید و برای حل آن و به دست آوردن ضرایب کنترل‌کننده از الگوریتم بهینه‌سازی محدب استفاده می‌شود. همچنین، مساله طراحی کنترل‌کننده مقاوم نیز درنظر گرفته شده است. در نهایت، یک مثال عددی و یک مثال عملی برای نشان دادن اعتبار و اثر بخشی روش طراحی شده ارائه می‌شود.

**کلید واژه-** برنامه‌ریزی خطی، تابع لیاپانوف خطی شبه مثبت، سیستم‌های فازی تاکاگی-سوگنو، سیستم‌های مثبت.

## 1- مقدمه

در بسیاری از سیستم‌های فیزیکی متغیرهای حالت معمولاً بیان کننده سطح جمعیت، دمای مطلق، غلظت و مقدار مواد هستند که ذاتاً شامل مقادیر غیرمنفی می‌باشند. این سیستم‌ها به طور معمول به عنوان سیستم‌های مثبت[1] شناخته می‌شوند [1]. سیستم‌های مثبت (یا به طور دقیق‌تر مثبت داخلی)، به سیستم هایی اطلاق می‌شود که خروجی و حالت‌ها، متناظر با هر حالت اولیه غیرمنفی و ورودی غیرمنفی، همیشه غیرمنفی باشند. چنین سیستم‌هایی تقریباً در تمامی شاخه‌های علم و تکنولوژی از جمله زیست‌شناسی و جمعیت‌شناسی، اقتصاد، مهندسی شیمی، پزشکی کاربرد دارند [2]. حالت‌های سیستم‌های مثبت به جای کل فضای خطی، درون یک مخروط واقع شده در ربع مثبت تعریف می‌شوند [3]. با توجه به ویژگی‌های منحصر به فرد و کاربردهای گسترده سیستم‌های مثبت، مسائل جدیدی در مورد آنها به وجود آمده و تجزیه و تحلیل این سیستم‌ها را به یک چالش تبدیل کرده است. اولین مرجع مهمی که به مطالعه سیستم‌های مثبت پرداخته و رویکرد نظری را برای آنها پیشنهاد می‌کند، توسط لئونبرگر در سال ۱۹۷۹ منتشر شد. از آن زمان به بعد سیستم‌های مثبت بسیار مورد توجه قرار گرفتند و مقالات متعددی به بررسی ویژگی‌ها و خصوصیات مهم این سیستم‌ها پرداخته‌اند [4-5]. بر خلاف سیستم‌های خطی، پایداری و پایدارسازی سیستم های مثبت با دو رویکرد مختلف مورد بررسی قرار می‌گیرند که اساساً به انتخاب نوع تابع لیاپانوف وابسته است، مساله طراحی کنترل‌کننده فیدبک حالت با رویکرد نامساوی ماتریسی خطی ($LMI^2$) در [6] و بر اساس برنامه‌ریزی خطی ($LP^3$) در [7] بررسی شده است. در میان تعداد انبوهی از تحقیقات انجام شده بر روی سیستم‌های مثبت [9-8]، به مسائل مربوط به سیستم‌های غیرخطی مثبت توجه کمتری شده است و تحقیقات نادری در این زمینه وجود دارد [10-12]. همین موضوع انگیزه‌ای برای کار حاضر بوده است. از آنجایی که مدل فازی تاکاگی-سوگنو ($T-S^4$) به عنوان یک تقریب زننده جامع در نظر گرفته می‌شود که می‌تواند یک تابع غیرخطی را با هر درجه‌ای از دقت تقریب بزند، برای نمایش معادلات غیرخطی سیستم‌های مثبت بسیار مناسب است.

با توجه به اینکه اکثر سیستم‌های فیزیکی و فرآیندهای عملی غیرخطی هستند و با تاکید بر این واقعیت که بسیاری از اهداف کنترلی به حل یک مساله پایداری منجر می‌شوند و همچنین توجه به این نکته که نتایج بدست آمده در سیستم‌های خطی مثبت را نمی‌توان برای سیستم‌های غیرخطی مثبت بکار برد

---

[1] Positive Systems

[2] Linear Matrix Inequality (LMI)

[3] Linear Programming (LP)

[4] Takagi-Sugeno (T-S)

[13]، مساله اصلی در طرح این مقاله، بررسی پایداری و پایدارسازی سیستم‌های مثبت با دینامیک غیرخطی، توصیف شده با مدل فازی تاکاگی-سوگنو با رویکردی جدید می‌باشد. برای اطمینان از پایداری و مثبت بودن سیستم حلقه بسته، بعضی قیود اضافی در مساله بهینه سازی معرفی می‌شوند و بر خلاف پژوهش‌های پیشین در حوزه سیستم‌های فازی T-S، از رویکرد برنامه‌ریزی خطی برای حل آنها استفاده می‌کنیم. با توجه به اینکه بسیاری از سیستم‌های عملی تحت تاثیر تغییرات محیطی، عدم قطعیت و آشفتگی‌ها هستند، مساله طراحی کنترل کننده مقاوم نیز در نظر گرفته شده است.

## 2- مقدمات و تعاریف اولیه

مدلسازی T-S روشی سیستماتیک است که در آن، معادلات غیرخطی سیستم اصلی توسط یک سری قواعد اگر-آنگاه به صورت چندین زیر سیستم خطی بیان می‌شوند و سپس با غیر فازی کردن آنها می‌توان دقیقا رفتار سیستم اصلی را مدل کرد. سیستم فازی توصیف شده توسط قانون $i$ ام زیر را درنظر بگیرید:

$R_i: \text{IF } z_1(k) \text{ is } M_{i1} \text{ and},...,\text{and } z_n(k) \text{ is } M_{in}, \text{THEN}$

$$\begin{cases} x(k+1) = A_i x(k) + B_i u(k), \\ y(k) = C_i x(k) + D_i u(k), \\ x(0) = x_0 \geq 0, \end{cases} \quad (1)$$

که در آن $x(k) \in \mathbb{R}^n$، $u(k) \in \mathbb{R}^m$ و $y(k) \in \mathbb{R}^t$ به ترتیب بردار حالت، ورودی کنترلی و بردار خروجی می‌باشند. $z_i$ ها متغیرهای فازی، $M_i$ ها مجموعه‌های فازی و $r$ تعداد قواعد می‌باشند. سپس، مدل کلی به صورت زیر توصیف می شود:

$$\begin{cases} x(k+1) = \sum_{i=1}^{r} h_i(z(k))(A_i x(k) + B_i u(k)), \\ y(t) = \sum_{i=1}^{r} h_i(z(k))(C_i x(k) + D_i u(k)), \\ x(0) = x_0 \geq 0, \end{cases} \quad (2)$$

که

$h_i(z(k)) = \frac{w_i(z(k))}{\sum_{i=1}^{r} w_i(z(k))}, \quad w_i(z(k)) = \prod_{j=1}^{\mu} M_{ij}(z_j(k)),$

$\sum_{i=1}^{r} h_i(z(k)) = 1, \quad 0 \leq h_i(z(k)) \leq 1, \quad i = 1,2,...,r$

$M_{ij}(z_j(k))$ درجه عضویت $x_j$ در $M_{ij}$ و $h_i(z(k))$ نشان دهنده توابع عضویت نرمالیزه شده می‌باشد.

در ادامه چندین لم و تعریف مهم در ارتباط با سیستم‌های مثبت معرفی می‌شوند.

**تعریف 2-1:**
سیستم فازی (2) یک سیستم مثبت کنترل‌شده گسسته-زمان نامیده می‌شود، اگر برای همه $x(0) \geq 0$ و $u(k) \geq 0$، همه مسیرهای متناظر برای $k \in \mathbb{N}$ در ربع مثبت باقی بماند.

**تعریف 2-2:**
ماتریس حقیقی $A$ شور نامیده می‌شود، اگر همه مقادیر ویژه آن اندازه‌ای کمتر از واحد داشته باشند و پایداری متناظر با آن پایدار شور نامیده می‌شود.

**لم 2-1 ([10]):**

سیستم گسسته-زمان (2) مثبت است، اگر و فقط اگر،

$A_i \ggeq 0, \ B_i \ggeq 0, \ C_i \ggeq 0, \ D_i \ggeq 0, \ i = 1,2,...,r.$

**لم 2-2 ([8]):**

فرض کنید $M^-$ و $M^+$ ماتریس‌های مربعی داده شده باشند. آنگاه، همه ماتریس‌ها در بازه $M^- \leq M \leq M^+$، پایدار شور و مثبت‌اند، اگر و فقط اگر، $M^-$ مثبت و $M^+$ پایدار شور باشد.

**نمادها:**

نماد $\mathbb{R}$ و $\mathbb{R}^n$ به ترتیب مجموعه اعداد حقیقی و فضای اقلیدسی $n$ بعدی را نشان می‌دهد. برای $x \in \mathbb{R}^n$، $x_i, i=1,...,n$ به معنی $i$ امین مختصه $x$ است. نماد $x >> 0$ ($x \ggeq 0$) به این معنی است که، برای $1 \leq i \leq n$، $x_i >> 0$ ($x_i \ggeq 0$) است. نماد $\|.\|$، نشان دهنده نرم اقلیدسی می باشد. $A^T$ بیانگر ماتریس ترانهاده و $\rho(A) = \max\{|\lambda_i(A)|, i=1,2,...,n\}$ شعاع طیفی ماتریس $A$ است.

## 3- نتایج تئوری:

در این بخش شرایط پایداری و پایدارسازی سیستم فازی (2) با استفاده از تابع لیاپانوف خطی شبه مثبت و رویکرد برنامه‌ریزی خطی مورد مطالعه قرار می‌گیرد.

### 3-1- شرایط پایداری

**قضیه 3-1:**

سیستم مثبت (2) با ورودی کنترلی $u(k) = 0$ پایدار مجانبی است اگر، بردار $p \ggeq 0$ (یا $p >> 0$) به نحوی وجود داشته باشد که حداقل یکی از دو مساله LP زیر برقرار شود:

LP1: $\quad p^T(A_i - I) << 0,$ \quad (3)

یا

$$\text{LP2:} \quad (A_i - I)p \ll 0. \tag{۴}$$

اثبات: تابع لیاپانوف $V(x(k)) = p^T x(k)$ را درنظر بگیرید. تفاضل مستقیم $V(x(k))$ به صورت زیر محاسبه می‌شود:

$$\Delta V(x(k)) = p^T x(k+1) - p^T x(k)$$
$$= p^T \sum_{i=1}^{r} h_i(z(k)) A_i x(k) - p^T x(k)$$
$$= \sum_{i=1}^{r} h_i(z(k)) p^T (A_i - I) x(k)$$

با توجه به $0 \leq h_i(z(k)) \leq 1$ و رابطه (۳)، می‌توان نتیجه گرفت که $\Delta V(x(k)) < 0$ است و اثبات کامل می‌شود.

یادآوری ۱: دوگان سیستم (۲) پایدار مجانبی است اگر و فقط اگر، سیستم (۲) پایدار مجانبی باشد. می‌توان با انتخاب تابع لیاپانوف $V(x(k)) = x^T(k) p$ برای سیستم دوگان نتیجه گرفت که رابطه (۴) برقرار است و معادل با رابطه (۳) می‌باشد.

### ۳-۲- طراحی کنترل کننده فیدبک حالت

در این زیربخش، با فرض اینکه تمامی متغیرهای حالت به طور کامل قابل اندازه گیری می‌باشند، هدف، طراحی قانون کنترلی فیدبک حالت با استفاده از ساختار کنترل‌کننده توزیع شده موازی به صورت زیر،

$$\Omega: u(k) = \sum_{i=1}^{r} h_i(z(k)) K_i x(k) \tag{۵}$$

به گونه‌ای می‌باشد که سیستم حلقه بسته زیر با اعمال کنترل کننده $\Omega$ در (۵)،

$$\begin{cases} x(k+1) = \sum_{i=1}^{r} \sum_{j=1}^{r} h_i(z(k)) h_j(z(k)) (A_i + B_i K_j) x(k) \\ y(k) = \sum_{i=1}^{r} \sum_{j=1}^{r} h_i(z(k)) h_j(z(k)) (C_i + D_i K_j) x(k) \end{cases} \tag{۶}$$

مثبت و پایدار مجانبی باشد.

قضیه ۳-۲:

برای سیستم فازی مثبت (۲)، کنترل‌کننده $\Omega$ در (۵) به گونه‌ای وجود دارد که سیستم حلقه بسته (۶) مثبت و پایدار مجانبی باشد، اگر بردار $p = [p_1, ..., p_n]^T \in \mathbb{R}^n$ و بردارهای $\xi_1^j, ..., \xi_n^j \in \mathbb{R}^m$ شرایط LP زیر را ارضا کنند:

$$\begin{cases} p > 0, \\ (A_i - I)p + B_i \sum_{t=1}^{n} \xi_t^j < 0, \quad i,j = 1,2,...,r, \\ a_{ht}^i p_h + b_h^i \xi_t^j \geq 0, \quad h,t = 1,2,...,n, \end{cases} \tag{۷}$$

سپس، بهره کنترل‌کننده، $K_j$، به صورت زیر محاسبه می‌شود:

$$K_j = [k_1^j, ..., k_n^j], \quad j = 1,2,...,r. \tag{۸}$$

که در آن $k_s = p_s^{-1} \xi_s^j$، $s = 1,2,...,n$ می‌باشد.

اثبات: پایداری: با استدلالی مشابه قضیه ۳-۱ و با جایگزین نمودن ماتریس حلقه بسته $A_{cl} = A_i + B_i K_j$ در معادله (۴) و همچنین مطابق با رابطه زیر،

$$K_j p = [k_1^j, ..., k_n^j] \begin{bmatrix} p_1 \\ ... \\ p_n \end{bmatrix} = \sum_{t=1}^{n} k_t^j p_t = \sum_{t=1}^{n} \xi_t^j$$

می‌توان نتیجه گرفت سیستم حلقه بسته پایدار است.

مثبت بودن: با توجه به لم ۲-۱، می‌توان نشان داد که مسیرهای شروع شده از هر شرط اولیه، در ربع مثبت باقی خواهند ماند.

$$(A_i + B_i K_j)_{ht} = a_{ht}^i + b_h^i \frac{\xi_t^j}{p_h} = a_{ht}^i p_h + b_h^i \xi_t^j \geq 0,$$

رابطه فوق برقرار خواهد بود، اگر و فقط اگر رابطه (۷) برقرار باشد. به این ترتیب اثبات کامل می‌شود.

یادآوری ۲: نتایج فوق را می توان به سیستم‌های T-S مثبت تعمیم داد، در این مورد، قانون کنترل فیدبک-حالت باید مثبت باشد، برای رسیدن به این هدف کافی است قید $\xi_t^j \geq 0$، $t = 1,...,n$ را به قیود قضیه ۳-۲ افزود.

نتیجه ۳-۱:

برای ماتریس‌های مثبت $A_i$ و $B_i$، سیستم حلقه بسته (۶) پایدار مجانبی و مثبت است، اگر مساله برنامه ریزی خطی زیر امکان پذیر باشد،

$$\begin{cases} (A_i - I)p + B_i \sum_{t=1}^{n} \xi_t^j < 0, \quad i,j = 1,2,...,r, \\ \xi_t^j \geq 0, \quad t = 1,2,...,n, \\ p > 0, \end{cases} \tag{۹}$$

در این شرایط، ماتریس بهره کنترل‌کننده از رابطه (۸) به دست می‌آید، که $p_s$ و $\xi_s$ از حل مساله LP فوق محاسبه می‌شوند.

### ۳-۳- پایدارسازی مقاوم

با توجه به اینکه در عمل گاهی پارامترهای سیستم ناشناخته‌اند یا اینکه دارای عدم قطعیت هستند، رویکرد پیشنهادی را می‌توان برای طراحی کنترل‌کننده مقاوم نیز تعمیم داد. فرض کنید سیستم (۲) دارای عدم قطعیت است و ماتریس $A$ به بازه

$[\underline{A},\overline{A}] : A \in \Theta$ تعلق دارد. $\underline{A}$ و $\overline{A}$ به ترتیب، کران‌های پایین و بالا شناخته شده ماتریس $A$ هستند.

قضیه ۳-۳:

برای سیستم فازی (۲)، کنترل‌کننده مقاوم (۵) به نحوی وجود دارد که سیستم حلقه بسته (۶) در سراسر بازه نامعین $\Theta$، مثبت و پایدار مقاوم باشد، اگر مساله LP زیر ارضا شود،

$$\begin{cases} p > 0, \\ (\overline{A}_i - I)p + B_i \sum_{t=1}^{n} \xi_t^j < 0, \ i,j = 1,2,...,r, \\ \underline{a}_{ht}^i p_h + b_h^i \xi_t^j \geq 0, \quad h,t = 1,2,...,n, \end{cases} \quad (10)$$

ماتریس بهره کنترل‌کننده از رابطه (۸) به دست می‌آید.

اثبات: با توجه به لم ۲-۲ و قضیه ۲-۳ اثبات دنبال می‌شود.

## ۴- نتایج شبیه سازی

در این بخش جهت نشان دادن اعتبار و اثربخشی رویکرد ارائه شده در این مقاله، یک مثال عددی و یک سیستم عملی مورد مطالعه قرار می‌گیرند.

مثال ۴-۱:

سیستم گسسته-زمان (۲) را با ماتریس‌های زیر درنظر بگیرید، [۲۷]:

$$A = \begin{bmatrix} 0.5 + 0.1\sin x_1(k) & 0.6 \\ 0.6 & 0.4 \end{bmatrix}, \ B = \begin{bmatrix} 0.1 \\ 0.2 \end{bmatrix},$$
$$C = \begin{bmatrix} 0.2 & 0.5 \end{bmatrix}, \ D = 0.1.$$

دو قانون فازی عبارت اند از:

Rule1) IF $z(k) = \sin x_1(k)$ is 1, THEN

$$\begin{cases} x(k+1) = A_1 x(k) + B_1 u(k), \\ y(k) = C_1 x(k) + D_1 u(k), \end{cases}$$

Rule 2) IF $z(k) = \sin x_1(k)$ is $-1$, THEN

$$\begin{cases} x(k+1) = A_2 x(k) + B_2 u(k), \\ y(k) = C_2 x(k) + D_2 u(k), \end{cases}$$

که در آن،

$$A_1 = \begin{bmatrix} 0.6 & 0.6 \\ 0.6 & 0.4 \end{bmatrix}, \ A_2 = \begin{bmatrix} 0.4 & 0.6 \\ 0.6 & 0.4 \end{bmatrix},$$
$$B_1 = B_2 = \begin{bmatrix} 0.1 \\ 0.2 \end{bmatrix}, \ C_1 = C_2 = \begin{bmatrix} 0.2 & 0.5 \end{bmatrix}, \ D_1 = D_2 = 0.1.$$

با در نظر گرفتن توابع عضویت به صورت زیر:

$$h_1(z(k)) = \frac{1 + \sin x_1(k)}{2}, \ h_2(z(k)) = 1 - h_1(z(k)),$$

مدل فازی سیستم غیرخطی عبارت است از:

$$\begin{cases} x(k+1) = \sum_{i=1}^{2} h_i(z(k))(A_i x(k) + B_i u(k)), \\ y(k) = \sum_{i=1}^{2} h_i(z(k))(C_i x(k) + D_i u(k)). \end{cases} \quad (11)$$

با حل شرایط برنامه‌ریزی خطی قضیه ۳-۲ در معادله (۷)، متغیرهای تصمیم‌گیری به صورت زیر محاسبه می‌شوند،

$$p = \begin{bmatrix} 152.6164 & 126.9691 \end{bmatrix}^T,$$
$$z = \begin{bmatrix} -110.8075 & -152.0953 & -110.8075 & -152.0953 \end{bmatrix},$$

با استفاده از رابطه (۸)، بهره کنترل‌کننده به صورت زیر به دست می‌آید:

$$K_1 = K_2 = \begin{bmatrix} -0.7261 & -1.1979 \end{bmatrix}.$$

ماتریس سیستم حلقه بسته عبارت اند از:

$$(A_1)_{cl} = \begin{bmatrix} 0.5274 & 0.4802 \\ 0.4548 & 0.1604 \end{bmatrix}, \ (A_2)_{cl} = \begin{bmatrix} 0.3274 & 0.4802 \\ 0.4548 & 0.1604 \end{bmatrix}.$$

به آسانی از نتایج بالا و لم ۱ می‌توان نتیجه گرفت که سیستم حلقه بسته مثبت می‌باشد. با محاسبه شعاع طیفی ماتریس $A_{cl}$،

$$\rho_1(A) = \begin{bmatrix} 0.8460 \\ 0.1581 \end{bmatrix}, \ \rho_2(A) = \begin{bmatrix} 0.7186 \\ 0.2308 \end{bmatrix}$$

به وضوح دیده می‌شود سیستم حلقه بسته پایدار شور است. شکل ۱، تغییرات متغیرهای حالت سیستم را در غیاب ورودی به ازای شرایط اولیه $x_0 = \begin{bmatrix} 0.01 & 0.03 \end{bmatrix}^T$، نشان می‌دهد. همان طور که دیده می‌شود سیستم حلقه باز ناپایدار است. با اعمال کنترل‌کننده، در شکل ۲، پاسخ حلقه بسته سیستم را می‌توان در شکل ۳ مشاهده نمود.

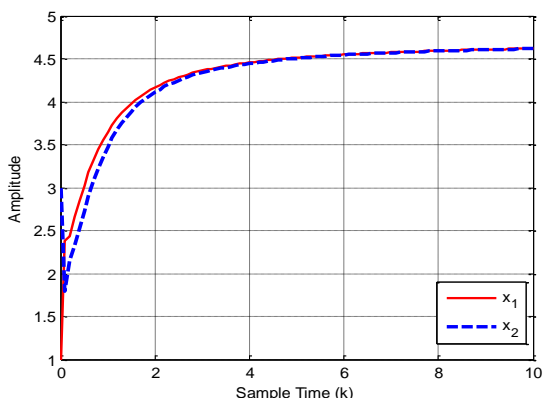

شکل ۱: پاسخ متغیرهای حالت سیستم در غیاب ورودی کنترلی

در این مثال به منظور نشان دادن محافظه کاری کمتر روش ارائه

شده در این مقاله، نسبت به کارهای پیشـین، پارامترهـای $a$ و $b$ برای مقایسه ناحیه پایداری به صورت زیر درنظر گرفته می‌شوند.

$$A_1 = \begin{bmatrix} 0.6 & 0.6 \\ a & 0.4 \end{bmatrix}, A_2 = \begin{bmatrix} 0.4 & 0.6 \\ b & 0.4 \end{bmatrix}, B_1 = B_2 = \begin{bmatrix} 0.1 \\ 0.2 \end{bmatrix}$$

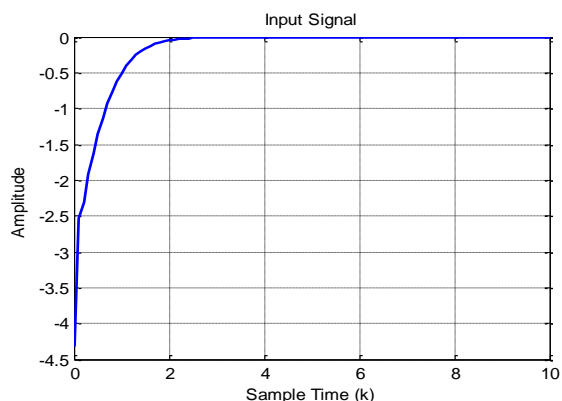

شکل ۲: ورودی کنترلی سیستم

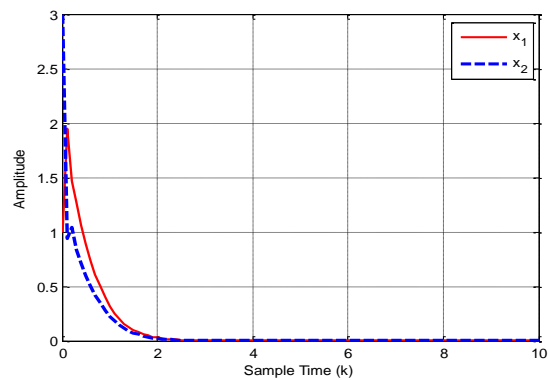

شکل ۳: پاسخ متغیرهای حالت سیستم حلقه بسته

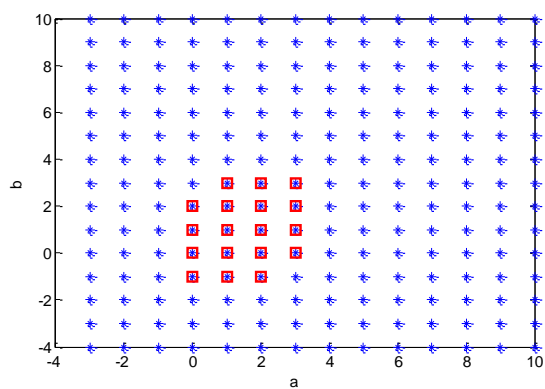

شکل ۴: ناحیه پایداری به ازای تغییرات $a$ و $b$

با توجه به شکل ۲ و ۳، می‌توان گفت که ورودی کنترلی به دست آمده یک ورودی کنترلی قابل قبول می‌باشد. همچنین، کنترل کننده طراحی شده به خوبی توانسته متغیرهای حالت سیستم را در مدت زمان محدود به پایداری برساند. شکل ۴، ناحیه پایداری را به ازای کنترل‌کننده (۵) و همچنین کنترل‌کننده ارئه شده در مرجع [۱۰] نشان می‌دهد. در این شکل، ٭ ناحیـه پایـداری را طبق کنترل‌کننده (۵) و □ ناحیه پایداری تحـت پوشـش طبـق کنترل‌کننده ارائه شده در مرجع [۱۰] را نشان می‌دهنـد. همـان طور که دیده می‌شود کنتـرل‌کننـده $\Omega$ در معادلـه (۵)، بـرای ناحیه گسترده‌ای از تغییـرات، پایـداری سیسـتم (۶) را تضمـین می‌کند.

مثال ۴-۲:

مثال عملی معمول سیستم‌هـای مثبـت، یعنـی مـدل دینـامیکی جمعیت ساختار آفت را مورد بررسی قرار می‌دهـیم [۱۲]. مـدل جمعیت انسانی یک سیستم مثبت است کـه در آن متغیرهـای حالت نشان دهنده تراکم و یا تعدادی از افراد جامعه است. چنین مدلی همیشه به صورت یک سیستم غیرخطی توصیف می‌شود. برای نشان دادن نتـایج تئـوری بـه دسـت آمـده، عـدم قطعیـت پارامتری که در آن $0 \leq \alpha \leq 1$، در این مدل درنظر گرفتـه شـده است. سیستم غیرخطی زیر درنظر بگیرید:

$$x(k+1) = A(k)x(k) + B(k)u(k)$$

کـه در آن $x(k) = [x_1(k), x_2(k), x_3(k)]^T$، $u(k) \in \mathbb{R}$ ورودی کنترلی و

$A(k) =$
$$\begin{bmatrix} 0.08 + 0.01\cos^2 x_1 - 0.1\alpha & 0.04 + 0.01\sin^2 x_1 + 0.1\alpha & 0.04 \\ 0.04 & 0 & 0 \\ 0 & 0.04 & 0 \end{bmatrix},$$

$$B(k) = \begin{bmatrix} 0.1 + 0.005\sin^2(x_1(k)) & 0 & 0 \end{bmatrix}^T,$$

بـا تعریـف تـرم غیرخطـی بـه صـورت $z(k) = \sin(x_1(k))$، $\max\{z(k)\} = 1$ و $\min\{z(k)\} = -1$، توابع عضویت بـه صـورت زیر به دست می‌آیند:

$$h_1(z(k)) = \frac{(z(k)+1)}{2}, \ h_2(z(k)) = \frac{(1-z(k))}{2}.$$

Rule 1: IF $z(k) = 1$, THEN: $x(k+1) = A_1 x(k) + B_1 u(k)$

Rule 2: IF $z(k) = -1$, THEN: $x(k+1) = A_2 x(k) + B_2 u(k)$

که در آن،

$$\bar{A}_1 = \begin{bmatrix} 0.09 & 0.04 & 0.04 \\ 0.04 & 0 & 0 \\ 0 & 0.04 & 0 \end{bmatrix}, \ \underline{A}_1 = \begin{bmatrix} -0.02 & 0.15 & 0.04 \\ 0.04 & 0 & 0 \\ 0 & 0.04 & 0 \end{bmatrix},$$

$$\bar{A}_2 = \begin{bmatrix} 0.08 & 0.05 & 0.04 \\ 0.04 & 0 & 0 \\ 0 & 0.04 & 0 \end{bmatrix}, \ \underline{A}_2 = \begin{bmatrix} -0.01 & 0.14 & 0.04 \\ 0.04 & 0 & 0 \\ 0 & 0.04 & 0 \end{bmatrix},$$

$B_1 = \begin{bmatrix} 0.105 & 0 & 0 \end{bmatrix}^T$ , $B_2 = \begin{bmatrix} 0.1 & 0 & 0 \end{bmatrix}^T$

با اعمال نتایج قضیه ۳-۳ در (۱۰)، بهره کنترل‌کننده با حل شرایط LP به دست آمده، به صورت زیر محاسبه می‌شود،

$K_1 = K_2 = \begin{bmatrix} 0.5399 & 0.5342 & 0.6753 \end{bmatrix}$

با اعمال کنترل‌کننده به سیستم، پاسخ حلقه بسته سیستم به ازای شرط اولیه $x(0) = \begin{bmatrix} 4.5688 & 1.4694 & 3.0119 \end{bmatrix}^T$ در شکل‌های ۵، ۶ و ۷ نشان داده شده است.

خوبی توانسته است در حضور عدم‌قطعیت، با تحمیل قید مثبت بودن سیستم حلقه بسته، متغیرهای حالت سیستم را در مدت زمان محدود به پایداری برساند.

## ۵- نتیجه‌گیری

در این مقاله مساله پایداری و پایدارسازی سیستم‌های فازی مثبت با رویکردی جدید مورد بررسی قرار گرفت. بر مبنای شرایط پایداری بدست آمده، شرایط لازم و کافی برای طراحی کنترل‌کننده که مثبت و پایدار بودن سیستم حلقه بسته را تضمین کند، در چارچوب مسائل برنامه‌ریزی خطی بیان شده است. همچنین، نشان داده شد این مساله حتی با وجود عدم قطعیت در پارامترهای سیستم، قابل حل می‌باشد. در نهایت، اعتبار و کارایی نتایج نظری به دست آمده با دو مثال نشان داده شد.

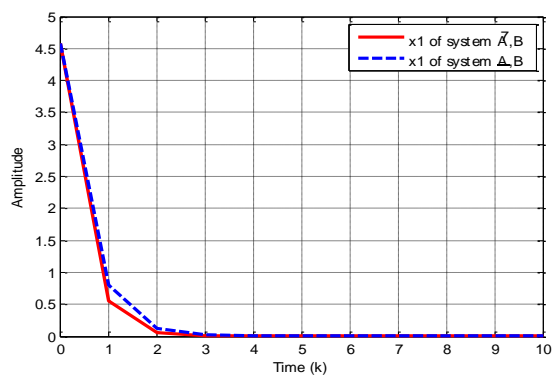

شکل ۵: پاسخ حلقه بسته متغیر حالت $x_1$

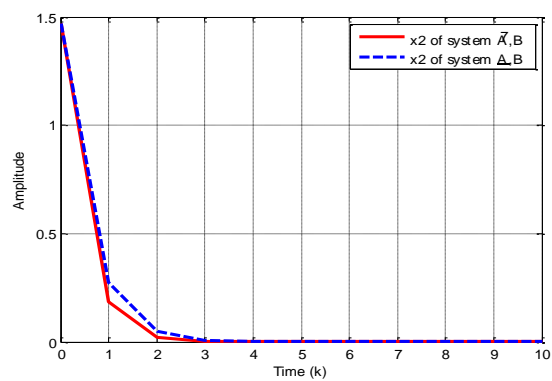

شکل ۶: پاسخ حلقه بسته متغیر حالت $x_2$

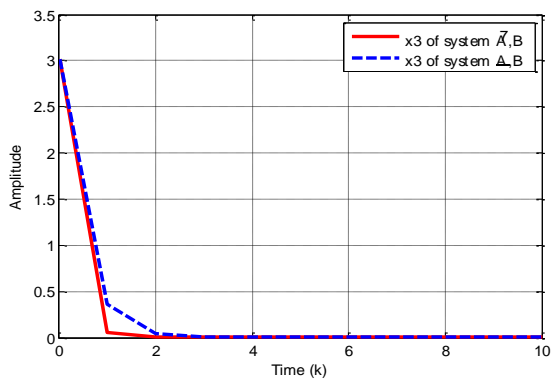

شکل ۷: پاسخ حلقه بسته متغیر حالت $x_3$

به وضوح دیده می‌شود که کنترل‌کننده مقاوم طراحی شده به